\documentstyle[prl,aps,graphicx,twocolumn]{revtex}
\begin{document}
\draft
\title{Quantum coherence and interaction--free measurements}
\author{Sierk P\"otting$^{1,2,3}$
        \thanks{email: sierk.poetting@wotan.opt-sci.arizona.edu},
        Eun Seong Lee$^1$,
        William Schmitt$^1$, Ilya Rumyantsev$^1$, Bernd Mohring$^1$ and Pierre Meystre$^1$}
\address{$^1$Optical Sciences Center, University of Arizona, Tucson,
Arizona 85721}
\address{$^2$Max--Planck--Institut f\"ur Quantenoptik, 85748 Garching, Germany}
\address{$^3$Sektion Physik, Universit\"at M\"unchen, 80333 M\"unchen, Germany}
\date{\today}
\maketitle
\begin{abstract}
We investigate the extent to which ``interaction-free''
measurements perturb the state of quantum systems.  We show that
the absence of energy exchange during the measurement is not a
sufficient criterion to preserve that state, as the quantum system
is subject to measurement dependent decoherence. While it is
possible in general to design interaction-free measurement schemes
that do preserve that state, the requirement of quantum coherence
preservation rapidly leads to a very low efficiency. Our results,
which have a simple interpretation in terms of ``which-way''
arguments, open up the way to novel quantum non-demolition
techniques.

\end{abstract}
 \narrowtext

Interaction-free measurements (IFM) are a particularly puzzling
example of the ``paradoxes'' illustrating the strangeness of quantum
physics. In a nutshell, they are measurements that offer a way to
detect the presence of an object without any apparent interaction
with the measuring device.

The history of IFM can be traced back to 1960, when Renninger
\cite{Renn60}, realized that at the quantum level, the
non-observation of a result does have a physical impact in that it
implies a collapse of the wave function. This point was further
investigated by Dicke \cite{Dick81} in the framework of
non-scattering of photons by particles. In 1993, Elitzur and 
Vaidman (EV) proposed a measurement scheme illustrating
particularly vividly the IFM paradox: They demonstrated that one
can to a certain extent determine the presence of a classical or
quantum mechanical object in an interferometer path without
touching it with a probe photon, and without prior information
about the location of the object. The EV scheme was soon
experimentally realized \cite{Kwia95} and its efficiency was
subsequently increased by use of the quantum Zeno effect
\cite{Kwia99}. Recent research has been directed towards
applications of the scheme, e.g. in the interaction-free imaging
of macroscopic objects with less than the classically expected
amount of light \cite{Whit98}.

Of course, IFMs are not really interaction-free: if the
interaction Hamiltonian between the object and the ``measuring
stick'', in most experimental cases a light beam, were to be set
equal to zero, then nothing would happen. But past this rather
trivial and easy way out of the problem, a more interesting
question is to try and quantify the meaning of interaction-free.
``Energy exchange free'' \cite{Pavi96} is now well established as
a more precise way to characterize IFMs in the case of classical
objects. This concept has also been applied at the quantum level,
where two--state systems have been investigated
\cite{Elit93,Karl98,Kwia96}. In a recent paper, White {\em et al}
\cite{Whit99a} showed that true IFM is not possible in the optical
detection of two-level atoms, due to the nonzero rate of forward
scattered photons. This gives a first indication that the quantum
mechanical situation is indeed more subtle than classical IFM.

It is well known that the quantum superpositions that oftentimes
characterize the state of an object are more sensitive to
interactions than its energy. In order to gain a full
understanding of IFMs, it is therefore important to analyze their
impact on such states. In addition, quantum superpositions
constitute a major ingredient of quantum entanglement. As such
they play a central role in quantum information processing,
including teleportation \cite{Benn93}, quantum computing
\cite{Shor94,Cira95} and cryptography \cite{Eker91,Rari94}.
Decoherence is one of the major obstacles to quantum information
processing, and the impact of IFMs on quantum coherence is an
important question in this context.

The goal of this paper is to analyze the impact of IFMs on quantum
coherence, as well as their interplay. Our main conclusion is that
even ``energy exchange free'' is an overly simplified description
of the situation. IFMs do generally change the state of the
system, as evidenced e.g. in the destruction of the quantum
superposition of two internal states as well as of the quantum
entanglement of two atoms. However, we also show that carefully
designed IFM schemes can provide a powerful and non-destructive
tool to probe quantum superpositions and entanglements. Potential
applications include the determination of the presence of ions in
a linear trap without disturbing their entanglement, and hence an
ongoing quantum computer calculation.

The situation that we consider is an extension of the EV scheme,
considering a multilevel atom in a coherent superposition of
electronic levels. Our goals are (a) to determine whether an IFM
can detect the presence of an atom without destroying that
superposition,  and (b) to quantify the impact of that
superposition on the outcome of the IFM measurements.

\begin{figure}
\begin{center}
\includegraphics[width=0.75\columnwidth]{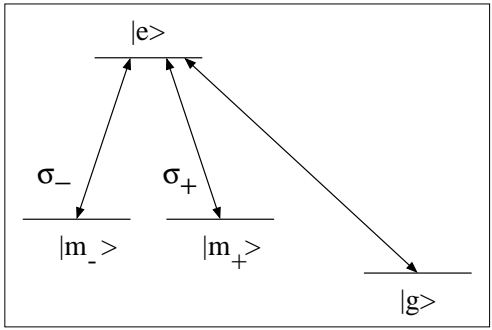}
\vspace{.3 cm}
 \caption{The atomic level scheme: The metastable states
  $|m_-\rangle$ and $|m_+\rangle$ are
  coupled to the excited state
  $|e\rangle$ via circularly polarized light. The state $|e \rangle$
  rapidly decays to a stable ground state $|g\rangle$
  which is far off-resonance from the
  metastable states. Once in this state, the atom is transparent
  to the light in the interferometer path.
  \label{figlevelscheme}
}
\end{center}
\end{figure}

An essential ingredient of the EV scheme is the fact that if the
photon path is along the arm of the interferometer where the
object is located, it will be irreversibly absorbed by this
object. Figure \ref{figlevelscheme} illustrates a model atomic
system that has this same property. It consists of a four-level
atom, initially in a general superposition of the two metastable
states $|m_-\rangle$ and $|m_+\rangle$.

\noindent The atoms can absorb single photons with unit
efficiency, inducing a transition to the excited level
$|e\rangle$, from which they can irreversibly decay to the ground
state $|g\rangle$. If the wavelength of the $|e\rangle
\longleftrightarrow |g\rangle$ transition is much larger than that
of the $|m_\pm\rangle \longleftrightarrow |e\rangle$ transitions,
then the branching ratio of these transitions is extremely high,
and reabsorption can be neglected. As a result, the absorption of
photons is de facto irreversible, the state $|g\rangle$ being
the``exploding bomb'' state of Ref. \cite{Elit93}. Filtering of
the high-frequency spontaneous photons circumvents the problem of
forward scattering described in Ref. \cite{Whit99a}.

\begin{figure}
\begin{center}
\includegraphics[width=0.75\columnwidth]{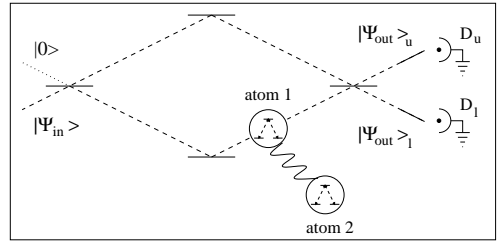}
\vspace{.3 cm} \caption{"Interaction-free" measurement on
  multi-level atoms.
  In addition to single atoms in internal superpositions one can also
  consider two entangled atoms, only one of them being in the
  interferometer path.
  \label{figinterferometer}
}
\end{center}
\end{figure}

As illustrated in Fig. \ref{figinterferometer}, the IFM scheme of
Ref. \cite{Elit93} is extended in such a way that the atom in the
lower arm of the interferometer is in a superposition of the
internal metastable states, the initial state of the atomic system
taken for concreteness to be
\begin{equation}
\label{eq:atomsuperposition}
|\phi_{atom}\rangle =
\frac{1}{\sqrt{2}} \left(|m_+\rangle+|m_-\rangle\right).
\end{equation}
This superposition can be interpreted as describing an atom which
is half absorbing and half transparent for a given photon
polarization. While at first sight similar to the superposition
states of being ``there'' or ``not there'' of Ref.
\cite{Elit93,Kwia96}, our ``half-absorbers'' are both located inside the
interferometer path and thus the polarization dependence of the two
transitions involved provides an additional degree of freedom. An
appropriate choice of field polarization enables us to
simultaneously probe both constituents of the superposition. This
opens up the way to additional control on the outcome on the IFM,
and in particular to the possibility of a non-demolition
measurement of the quantum superposition of Eq.
(\ref{eq:atomsuperposition}). As such, it permits the IFM of a
``quantum bomb'' in a superposition of its ``armed'' and
``unarmed'' states.

We denote the photon creation and annihilation operators by
$\hat{a}^\dagger_{k,\mu}$ and $\hat{a}_{k,\mu}$, respectively,
where $k=\{u,l\}$ stands for photons of polarization $\mu$
following the upper or lower path in Fig. \ref{figinterferometer}.
Hence, the ket $\hat{a}_{l,\mu}^\dagger|0\rangle$ describes a
photon of polarization $\mu$ in the lower interferometer arm. We
consider both circular polarizations $\sigma_\pm$, in which case
$\mu =\{+,-\}$, and linear polarizations $\mu=\{x,y\}$, the atomic
selection rules being such that the $|m_-\rangle \leftrightarrow
|e\rangle$ and $|m_+\rangle \leftrightarrow |e\rangle$ are excited
by left and right-circularly polarized light $\sigma_-$ and
$\sigma_+$, respectively. The absorption of a photon by the atom
can therefore be described by the transition
\begin{equation}
\label{eq:absorption}
\hat{a}_{l,\pm}^\dagger|0\rangle|m_\mu \rangle
\rightarrow
\left\{
\begin{array}
{r@{\quad {\mbox {for}} \quad}l} |{\cal S}\rangle|g \rangle & \mu
= \pm,
\\ \hat{a}_{l,\pm}^\dagger|0\rangle|m_\mp \rangle & \mu = \mp ,
\end{array}
\right.
\end{equation}
where $|{\cal S}\rangle$ is a scattered photon. We assume that
this high-frequency photon escapes the system without possible
reabsorption, and can be additionally be filtered away from the
detectors.

As in the original EV setup, a single photon enters the
interferometer via the lower input port in the state
\begin{equation}
|\psi_{light,in}\rangle = a^\dagger_{l,\mu}|0\rangle .
\end{equation}
Two identical non-polarizing 50-50 beam splitters constitute the
input and output of the interferometer. The phase shifts upon
transmission are chosen so that the upper detector $D_u$ fires
with unit probability in case if no atom is in the interferometer.
In that case, the annihilation operators $\hat{a}_{u,l}$ before and
$\hat{a}_{u,l}'$ after each beam splitter are related by
\begin{equation}
\label{eq:bsconcrete}
\left(\begin{array}{c} \hat{a}_u'\\ \hat{a}_l'\end{array}\right)
=
\frac{1}{\sqrt{2}}
\left(\begin{array}{cc}
          i & 1 \\
          1 & i
\end{array}\right)
\left(\begin{array}{c} \hat{a}_u \\ \hat{a}_l \end{array}\right),
\end{equation}
independently of the photon polarization \cite{Mandel}.

Consider first the case of a $\sigma_+$-polarized photon entering
the lower port of the first beam splitter, so that the initial
state of the atom-field system is
\begin{equation}
|\psi_{in}\rangle
=
\hat{a}_{l,+}^\dagger |0\rangle |\phi_{atom}\rangle .
\end{equation}

As the photon propagates through the interferometer, Eqs.
(\ref{eq:absorption}) and (\ref{eq:bsconcrete}) show that the
system evolves to the final state
\begin{eqnarray}
|\psi_{out}\rangle &=& -\frac{1}{2} |{\cal S}\rangle |g \rangle +
\frac{i}{2\sqrt{2}} \hat{a}_{l,+}^\dagger|0\rangle |m_+\rangle
\nonumber \\ &-& \frac{1}{2\sqrt{2}}
\hat{a}_{u,+}^\dagger|0\rangle |m_+\rangle  - \frac{1}{\sqrt{2}}
\hat{a}_{u,+}^\dagger|0\rangle |m_-\rangle. \label{psiout}
\end{eqnarray}
Since the interferometer tuning is such that the upper detector
$D_u$ clicks with unit probability in case no atom is present, the
register of a click on the lower detector indicates with certainty
the presence of an atom in the lower arm. As a result of the form
of the interaction between the atom and the light field, this
click can be interpreted as resulting from a ``photon propagating
through the upper arm'', hence an IFM. Eq. (\ref{psiout}) shows
that the probability for such an event is 1/8, a factor of 2 less
than in the classical absorber and two-level atom cases. This is
because in the present case, the atom can be thought of as a
``half-absorber'', a point to which we shall return shortly. After
detection of a click on $D_l$, the normalized reduced atomic
density operator is
\begin{equation}
\label{eq:traceplus} \rho_{atom, out} \propto \mbox{Tr}_{light}
\left\{ |\psi_{out}\rangle\langle\psi_{out}|
\hat{a}_{l,+}^\dagger\hat{a}_{l,+}^{} \right\} =
|m_+\rangle\langle m_+|.
\end{equation}
Hence, the IFM destroys the initial quantum superposition of the
atomic state, leaving it in the energy eigenstate $|m_+\rangle$.

This results from the fact that an atom initially in state
$|m_-\rangle$ is transparent to $\sigma_+$-polarized light. Hence,
if a signal is registered at detector $D_l$ we know for sure that
the atom had to be initially in state $|m_+\rangle$. A measurement
scheme using circularly polarized light therefore provides ``which
path'' information about the atomic state, and leads to a
projection of its state onto $|m_+\rangle$. Likewise, using
$\sigma_-$-polarized light projects the superposition  to
$|m_-\rangle$ upon detection with $D_l$.

The ``which path'' information can be erased by using linearly
polarized light, say along the $x$-direction. Decomposing this
polarization into its circular components, the initial atom-field
system is now in the state
\begin{equation}
|\psi_{\mathit{in}}\rangle
=
\frac{1}{\sqrt{2}}
\left(
\hat{a}_{l,-}^\dagger - \hat{a}_{l,+}^\dagger
\right)|0\rangle
|\phi_{\mathit{atom}}\rangle,
\end{equation}
and yields the final state
\begin{eqnarray}
\label{eq:afterlinear}
|\psi_{\mathit{out}}\rangle &=& -\frac{1}{2\sqrt{2}} \left( |{\cal
S}\rangle|g\rangle-|{\cal S}'\rangle|g\rangle \right) \nonumber\\
&+& \frac{1}{4} \left( \hat{a}_{u,+}^\dagger|0\rangle|m_+\rangle -
\hat{a}_{u,-}^\dagger|0\rangle|m_-\rangle \right) \nonumber \\ &+&
\frac{1}{2} \left( \hat{a}_{u,+}^\dagger|0\rangle|m_-\rangle -
\hat{a}_{u,-}^\dagger|0\rangle|m_+\rangle \right) \nonumber \\
 &-& \frac{i}{4} \left(
\hat{a}_{l,+}^\dagger|0\rangle|m_+\rangle -
\hat{a}_{l,-}^\dagger|0\rangle|m_-\rangle \right).
\end{eqnarray}
It is the last term of this expression, the maximally entangled
atom-photon state
\begin{equation}
|\psi_{\mathit{out}} \rangle_l 
\propto
\hat{a}_{l,+}^\dagger|0\rangle|m_+\rangle -
\hat{a}_{l,-}^\dagger|0\rangle|m_-\rangle , \label{eq:detectorlow}
\end{equation}
which is of interest to us, since it is associated with the
detection of light on the lower detector $D_l$. Karlsson and
coworkers \cite{Karl98} discussed a similar state in the case of
the IFM detection of a two-level atom and proposed its use to make
a nondemolition measurement of the ground--state atom number. The
present situation is different in that the entanglement is now in
the state of the coupled atom-field system associated with the
output at just one arm of the interferometer. As we now show, this
entanglement can be used to perform an interaction-free, quantum
non-demolition measurement of the quantum superposition of the
atomic state. More generally, we can make use of the mapping of
the atomic and photon states associated with the entanglement to
encode an atomic superposition in the state of the measured
photon. The associated selective measurement then leaves the atom
in the desired quantum superposition.

We proceed by re-expressing the normalized state
$|\Psi_{\mathit{out}}\rangle_l$ in terms of linearly polarized
light components as
\begin{eqnarray}
\label{eq:xypolarized} |\psi_{\mathit{out}}\rangle_l &=&
\frac{i}{\sqrt{2}} \hat{a}_{l,x}^\dagger|0\rangle \left(
|m_+\rangle + |m_-\rangle \right) \nonumber \\ &-&
\frac{1}{\sqrt{2}} \hat{a}_{l,y}^\dagger|0\rangle \left(
|m_+\rangle - |m_-\rangle \right).
\end{eqnarray}
From this result it is immediately apparent that a
polarization-sensitive measurement of $x$-polarized photons leaves
the atoms in the pure final state
\begin{equation}
\label{eq:xpolarized} |\phi_{atom}\rangle_{out} = \frac{1}{\sqrt
2} (|m_+\rangle + |m_-\rangle),
\end{equation}
which is precisely the initial superposition
$|\phi_{atom}\rangle$, while a detection of $y$-polarized photons
gives the orthogonal final state
\begin{equation}
|\phi_{atom}\rangle_{out} = \frac{1}{\sqrt 2} (|m_+\rangle -
|m_-\rangle).
\end{equation}
In each case, the detection probability is readily seen from Eq.
(\ref{eq:afterlinear}) to be 1/16, a factor of 2 less than
previously and a factor of 4 less than in the case of two-level
atoms. We are able to preserve the initial superposition because
each circular polarized component performs an IFM on a different
``half-absorber'', located in the same interferometer path. The
appropriate measurement of a linear polarization, a superposition of 
both circular polarizations, then combines both results.  

It is important to remark that in the case of
polarization-insensitive detection, the final atomic state is not
a pure quantum superposition, but rather the mixture
\begin{equation}
\rho_{atom, out} = \frac{1}{2} \left (|m_+\rangle \langle m_+| +
|m_-\rangle \langle m_-| \right ).
\end{equation}

We see, then, that the atom-photon entanglement between of the
state $|\psi_{out}\rangle_l$ provides us with a tool not just to
perform an interaction-free, quantum-nondemolition measurement of
the state of an atom, but also to map its quantum coherence to a
prescribed value.

We already mentioned that compared to the original EV scheme
\cite{Elit93}, the probability of detecting the presence of an
atom in an interaction-free fashion, either without destroying its
initial state or preparing it in a prescribed superposition, is
considerably smaller than for two-level atoms. For the specific
example considered here, the reduction is by a factor of 4. This
results from the multiplicity of atomic and light polarization
states, i.e. the larger dimensionality of the
relevant Hilbert space. Of all the ``branches'' followed by the
wave function of the system during its evolution, only a few are
useful to reconstruct the initial superposition state. The
situation rapidly worsens for larger systems, and it is quite
clear that decoherence--free IFMs soon become unrealistic. We note
in addition that unlike in the EV scheme, there is no way to
guarantee the preservation of the atomic state in case the photon
is detected by the upper branch detector $D_u$. It is easily shown
that in that case, the post-measurement probability to find the
atom in its initial state is always less than unity, due to the fact
that there are two orthogonal maximally entangled contributions in 
Eq. (\ref{eq:afterlinear}) associated
with the upper detector. Consequently,
the initial state of the object has to be reset before a
subsequent measurement can be performed, in contrast to the EV
situation.

Despite these difficulties, it should be emphasized that the
polarization-dependent measurement scheme presents advantages of
considerable interest for quantum information processing
applications. In particular, it is easily extended to the
situation of entangled atoms, as illustrated in Fig.
\ref{figinterferometer}. Assume for concreteness that the two
particles are initially in the Bell state
$$\frac{1}{\sqrt{2}}\left(|m_-\rangle_1|m_+\rangle_2 +
|m_+\rangle_1|m_-\rangle_2\right), $$ where $|\cdot\rangle_i$
corresponds to atom $i$. As a result of the local character of the
measurement scheme, it follows that a choice of
polarization-sensitive detection that preserves the quantum
coherence of the atom inside the interferometer also preserves its
entanglement with the other atom, while decoherence also implies
the destruction of the entanglement. As such, decoherence--free
IFMs provide a tool to monitor the presence of atoms without
destroying their state of entanglement.

In summary, it is possible to determine the presence of an atom in
a quantum superposition of internal states without destroying it,
provided that the measurement scheme does not provide the ``which
way'' information that would in principle permit to determine its
internal state. This implies that when aimed at measuring
multi-level atoms, IFMs have to be designed exceedingly carefully.
Compared to the classical case their efficiency is very low,  and
information is lost even if the upper detector detects a photon.
We also showed how polarization-sensitive IFMs can be used to map
the polarization state of the detected photon onto the internal
state of the atoms, and how these measurements translate directly
to the domain of entangled atoms. Although our scheme does not
actively prevent a system from decoherence, it opens up
possibilities to better control quantum systems and monitor
quantum systems, with potential applications in quantum
information processing.

We wish to thank M.~G. Moore and M.~Wilkens for stimulating
discussions. This work was supported in part by the Office of Naval Research
Research Contract No. 14-91-J1205, the National Science Foundation
Grant PHY98-01099, the US Army Research Office and the Joint
Services Optics Program.

\end{document}